\documentclass[%
 reprint,
 amsmath,amssymb,
 aps,
]{revtex4-2}

\usepackage[T1]{fontenc}

\usepackage{graphicx}
\usepackage{dcolumn}
\usepackage{bm}

\begin{document}

\preprint{APS/123-QED}

\title{Nuclear Spin-Depleted, Isotopically Enriched $^\text{70}$Ge/$^\text{28}$Si$^\text{70}$Ge Quantum Wells}

\author{O. Moutanabbir}
\email{oussama.moutanabbir@polymtl.ca}
\affiliation{Department of Engineering Physics, \'Ecole Polytechnique de Montr\'eal, Montr\'eal, C.P. 6079, Succ. Centre-Ville, Montr\'eal, Qu\'ebec, Canada H3C 3A7}
\author{S. Assali}
\affiliation{Department of Engineering Physics, \'Ecole Polytechnique de Montr\'eal, Montr\'eal, C.P. 6079, Succ. Centre-Ville, Montr\'eal, Qu\'ebec, Canada H3C 3A7}
\author{A. Attiaoui}
\affiliation{Department of Engineering Physics, \'Ecole Polytechnique de Montr\'eal, Montr\'eal, C.P. 6079, Succ. Centre-Ville, Montr\'eal, Qu\'ebec, Canada H3C 3A7}
\author{G. Daligou}
\affiliation{Department of Engineering Physics, \'Ecole Polytechnique de Montr\'eal, Montr\'eal, C.P. 6079, Succ. Centre-Ville, Montr\'eal, Qu\'ebec, Canada H3C 3A7}
\author{P. Daoust}
\affiliation{Department of Engineering Physics, \'Ecole Polytechnique de Montr\'eal, Montr\'eal, C.P. 6079, Succ. Centre-Ville, Montr\'eal, Qu\'ebec, Canada H3C 3A7}
\author{P. Del Vecchio}
\affiliation{Department of Engineering Physics, \'Ecole Polytechnique de Montr\'eal, Montr\'eal, C.P. 6079, Succ. Centre-Ville, Montr\'eal, Qu\'ebec, Canada H3C 3A7}
\author{S. Koelling}
\affiliation{Department of Engineering Physics, \'Ecole Polytechnique de Montr\'eal, Montr\'eal, C.P. 6079, Succ. Centre-Ville, Montr\'eal, Qu\'ebec, Canada H3C 3A7}
\author{L. Luo}
\affiliation{Department of Engineering Physics, \'Ecole Polytechnique de Montr\'eal, Montr\'eal, C.P. 6079, Succ. Centre-Ville, Montr\'eal, Qu\'ebec, Canada H3C 3A7}
\author{N. Rotaru}
\affiliation{Department of Engineering Physics, \'Ecole Polytechnique de Montr\'eal, Montr\'eal, C.P. 6079, Succ. Centre-Ville, Montr\'eal, Qu\'ebec, Canada H3C 3A7}

\date{\today}

\begin{abstract}
The p-symmetry of the hole wavefunction is associated with a weaker hyperfine interaction as compared to electrons, thus making hole spin qubits attractive candidates to implement long coherence quantum processors. However, recent studies demonstrated that hole qubits in planar germanium (Ge) heterostructures are still very sensitive to nuclear spin bath. These observations highlight the need to develop nuclear spin-free Ge qubits to suppress this decoherence channel and evaluate its impact. With this perspective, this work demonstrates the epitaxial growth of $^\text{73}$Ge-depleted isotopically enriched $^\text{70}$Ge/SiGe quantum wells. The growth was achieved by reduced pressure chemical vapor deposition using isotopically purified monogermane $^\text{70}$GeH$_\text{4}$ and monosilane $^\text{28}$SiH$_\text{4}$ with an isotopic purity higher than 99.9 $\%$ and 99.99 $\%$, respectively. The quantum wells consist of a series of $^\text{70}$Ge/SiGe heterostructures grown on Si wafers using a Ge virtual substrate and a graded SiGe buffer layer. The isotopic purity is investigated using atom probe tomography following an analytical procedure addressing the discrepancies in the isotopic content caused by the overlap of isotope peaks in mass spectra. The nuclear spin background in the quantum wells was found to be sensitive to the growth conditions. The lowest concentration of nuclear spin-full isotopes $^\text{73}$Ge and $^\text{29}$Si in the heterostructure was established at 0.01 $\%$ in the Ge quantum well and SiGe barriers. The measured average distance between nuclear spins reaches 3-4 nm in $^\text{70}$Ge/$^\text{28}$Si$^\text{70}$Ge, which is an order of magnitude larger than in natural Ge/SiGe heterostructures.  

\end{abstract}

\maketitle


\section{\label{sec:level1}Introduction}

Although it was quickly relegated behind silicon (Si) because of its relatively low bandgap energy, its lack of a stable oxide, and its large surface state densities, germanium (Ge) is inarguably the material that catalyzed the transition from what W. Pauli and I. Rabi called the ‘’Physics of Dirt’’~\cite{anderson_1993, Cahn_2002} to modern-day semiconductor physics and technology ~\cite{HALLER2006408}. Indeed, the ease by which Ge can then be purified and processed led to the demonstration of point contact diode mixers for radar reception ~\cite{Torrey1948} and of the point contact and junction transistors ~\cite{Shockley1956}. These inventions contributed to laying the groundwork for what was later coined as the first quantum revolution. In recent years, there has been a revived interest in Ge-based materials for integrated photonic circuits ~\cite{MarrisMoriniVakarinRamirezLiuBallabioFrigerioMontesinosAlonsoRamosLeRouxSernaBenedikovicChrastinaVivienIsella+2018+1781+1793,Moutanabbir-APL2021}, sensing ~\cite{MarrisMoriniVakarinRamirezLiuBallabioFrigerioMontesinosAlonsoRamosLeRouxSernaBenedikovicChrastinaVivienIsella+2018+1781+1793}, high-mobility electronic s~\cite{Toriumi_2018}, and solid-state quantum computing ~\cite{Scappucci2020}. The latter, for instance, aims at capitalizing on the advantageous quantum environment of holes in Ge, their inherently large and tunable spin-orbit interaction (SOI), and their reduced hyperfine coupling with nuclear spins to implement increasingly robust and reliable spin qubits ~\cite{Hendrickx2020_1,Hendrickx2020_2,Hendrickx2021,Bogan2019,Wang2016,Watzinger2018,Jirovec2022,Wang2022,Bosco2021,Delvecchio2023}. Indeed, these quantum devices are now considered forefront candidates for scalable quantum processors ~\cite{Scappucci2020}. This recent surge in developing Ge qubits makes one think that Ge may also be a key material in shaping the anticipated second quantum revolution.  

From a fundamental standpoint, it is expected that the hyperfine interaction to be weaker for holes than electrons due to the p-symmetry of the hole wavefunction. However, theoretical investigations suggested a hyperfine coupling that is only one order of magnitude smaller than that of electrons ~\cite{Testelin2009,Philippopoulos2020} or of equal strength as in Si ~\cite{Philippopoulos2020}. Moreover, the p-symmetry and d-orbital hybridization of the hole wavefunction leads to an anisotropic hyperfine coupling that is non-existent for electron spins ~\cite{Testelin2009}. Interestingly, recent experimental studies hint at the sensitivity of hole spin qubits in planar Ge/SiGe heterostructure to nuclear spin bath reporting an amplitude of the fluctuating Overhauser field of 34.4 kHz, which is suggested to limit spin dephasing times ~\cite{LawrieAPS2022}. Although charge noise is believed to be the dominant decohering process, these observations call for the development of nuclear spin-free Ge qubits to elucidate their sensitivity to hyperfine coupling. Undertaking this research direction requires Ge-based quantum devices that are depleted of $^\text{73}$Ge, which is the only Ge nuclear spin-full stable isotope. This work addresses this very issue and provides a demonstration of the epitaxial growth of isotopically purified $^\text{70}$Ge quantum wells (QWs). Note that enriched $^\text{70}$Ge, $^\text{74}$Ge, and $^\text{76}$Ge isotopes were employed in the past to grow superlattices and self-assembled quantum dots by solid-source molecular beam epitaxy ~\cite{Fuchs1995,MiyamotoAB2010,Moutanabbir2010PRL}. Herein, the growth of $^\text{73}$Ge-depleted QWs is achieved by hydride precursors using the chemical vapor deposition (CVD) method, which is broadly adopted in Ge device research besides being compatible with the processing standards in the semiconductor industry ~\cite{Scappucci2020}.

\begin{figure*}[htb]
    \centering
    \includegraphics[width=1 \textwidth]{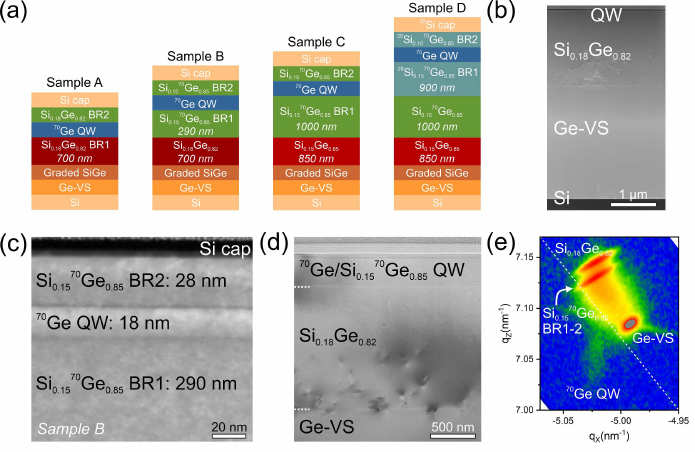}
    \caption{\textbf{Epitaxial growth of isotopically engineered Ge/SiGe heterostructures}. (a) Schematic illustration of four sets of the grown Ge/SiGe heterostructures. (b) A cross-sectional STEM image of an as-grown Ge/SiGe QW showing the entire stack consisting of Ge-Vs and SiGe buffer layers grown on a Si substrate. (c) A close-up cross-sectional STEM image of an isotopically enriched $^\text{70}$Ge/SiGe QW. (d) A cross-sectional TEM image showing that the QW region is defect-free and that the extended defects are confined at Ge-VS/SiGe interface. (e) The (224) XRD-RSM of a representative CVD-grown isotopically purified $^\text{70}$Ge QW heterostructure.}
    \end{figure*}

\section{\label{sec:level1}Experimental}

The epitaxial growth of isotopically engineered Ge/SiGe QW heterostructures was carried out on hydrogen-passivated 4-inch (001)-oriented Si wafers in a reduced-pressure CVD reactor using isotopically purified monogermane $^\text{70}$GeH$_\text{4}$ (isotopic purity $>$99.9 $\%$) and monosilane $^\text{28}$SiH$_\text{4}$ (isotopic purity $>$99.99 $\%$). The precursors were enriched in a centrifugal setup using natural monogermane ($^\text{nat}$GeH$_\text{4}$) and SiF$_\text{4}$ as starting gases ~\cite{Sozin2022}. After purification, $^\text{70}$GeH$_\text{4}$ contains traces ($<$0.006 at.$\%$) of other Ge isotopes: $^\text{72}$Ge, $^\text{73}$Ge, $^\text{74}$Ge, and $^\text{76}$Ge. Moreover, chemical contaminants including other hydrides are also negligible, with an average content being $<$0.06 µmol/mol. Reference Ge/SiGe QW heterostructures were also prepared following the same growth protocol using conventional precursors with natural isotopic abundance ($^\text{nat}$GeH$_\text{4}$ and disilane $^\text{nat}$Si$_\text{2}$H$_\text{6}$). After annealing in hydrogen, a ~3 µm-thick Ge interlayer, commonly known as Ge virtual substrate (Ge-VS), was grown on Si using $^\text{nat}$GeH$_\text{4}$ and a two-step growth process in the 450-600 °C temperature range. Then follows  a thermal cyclic annealing step (725-875 °C) to improve the Ge-VS quality. A reverse-graded ~1µm-thick Si$_\text{1-x}$Ge$_\text{x}$ layer was then grown at 600 °C using $^\text{nat}$GeH$_\text{4}$ and $^\text{nat}$Si$_\text{2}$H$_\text{6}$ until a uniform Si content of 18 at.$\%$ is reached. Without interrupting the growth, the $^\text{nat}$GeH$_\text{4}$ supply was switched to the purified $^\text{70}$GeH$_\text{4}$ to grow the first Si$_\text{1-x}$Ge$_\text{x}$ barrier layer (BR1), while keeping all the other growth parameters unchanged. Thickness and composition of BR1 were varied in the ~0.3-1µm range and x = 0.15-0.18 range to investigate the effect of the growth time on the isotopic purity of the epilayers. After that the growth of BR1 was completed, the reactor was then purged in hydrogen for 90 s before growing the $^\text{70}$Ge QW layer using $^\text{70}$GeH$_\text{4}$ supply for a variable growth time of up to 40 s. Next, the reactor was purged in hydrogen for 90 s prior to the growth of the Si$_\text{1-x}$Ge$_\text{x}$ BR2 layer under identical growth conditions as BR1. Lastly, a Si capping layer with a few nm thickness was grown. Fig. 1a illustrates the grown stacks. Ge/Si$_\text{0.18}$Ge$_\text{0.82}$ (A), $^\text{70}$Ge/Si$_\text{0.18}$Ge$_\text{0.82}$ (B), and $^\text{70}$Ge/Si$_\text{0.15}$Ge$_\text{0.85}$ (C) QWs were grown using this protocol. The $^\text{70}$Ge/$^\text{28}$Si$_\text{0.15}$$^\text{70}$Ge$_\text{0.85}$ (D) QW was grown following a similar protocol except for the growth of BR1-2 that was performed by changing from $^\text{nat}$Si$_\text{2}$H$_\text{6}$ to $^\text{28}$SiH$_\text{4}$ and adjusting the growth conditions to accommodate the change in the precursor decomposition.

Several characterization techniques were employed to elucidate the basic properties of the as-grown heterostructures and investigate their isotopic content. Lattice strain and average content in Ge/SiGe heterostructures were evaluated from X-ray diffraction (XRD) measurements including reciprocal space map (RSM) analysis. The microstructure of the grown materials was investigated by transmission electron microscopy (TEM) and scanning TEM (STEM). The quality of interfaces, the atomic-level composition, and the isotopic purity were investigated using atom probe tomography (APT). Additional insights into the chemical and isotopic compositions are also obtained using secondary ion mass spectrometry (SIMS). Raman scattering spectroscopy was employed to evaluate the effects of the isotopic content on phonon scattering in Ge QWs. Additionally, the uniformity of the growth thickness as well as the optical signature of quantum confinement were investigated using spectroscopic ellipsometry (SE).

\section{\label{sec:level1}Results and Discussion}

A  cross-sectional STEM image of a representative isotopically-engineered Ge QW heterostructures is shown in Fig. 1b, while the enlarged view of the $^\text{70}$Ge/Si$_\text{0.15}$$^\text{70}$Ge$_\text{0.85}$ QW region is displayed in Fig. 1c. The figure shows an 18 nm-thick $^\text{70}$Ge QW together with BR1 and BR2 layers with thicknesses of 290 nm and 28 nm, respectively. The transition between SiGe barrier layers and $^\text{70}$Ge QW is of the order of 1-2 nm. To evaluate the structural quality of the heterostructures, cross-sectional TEM images were acquired (Fig. 1d). The extended defects are confined to the Si/Ge-VS and Ge-VS/Si$_\text{1-x}$Ge$_\text{x}$ interfaces, with no defects being detected in the QW region at the TEM imaging scale. XRD-RSM (224) analysis of the as-grown heterostructures demonstrates sharp peaks for the SiGe/Ge substrate, barriers as well as the signature of the strained 18 nm-thick $^\text{70}$Ge layer, thus suggesting an excellent degree of crystallinity across the structure (Fig. 1(e)). Here, the variation in composition between natural and purified SiGe layers (Si$_\text{0.18}$Ge$_\text{0.82}$ vs. Si$_\text{0.15}$Ge$_\text{0.85}$ as determined by APT) is related to the difference in composition between the germane precursor supplies.

\begin{figure*}[htb]
    \centering
    \includegraphics[width=1 \textwidth]{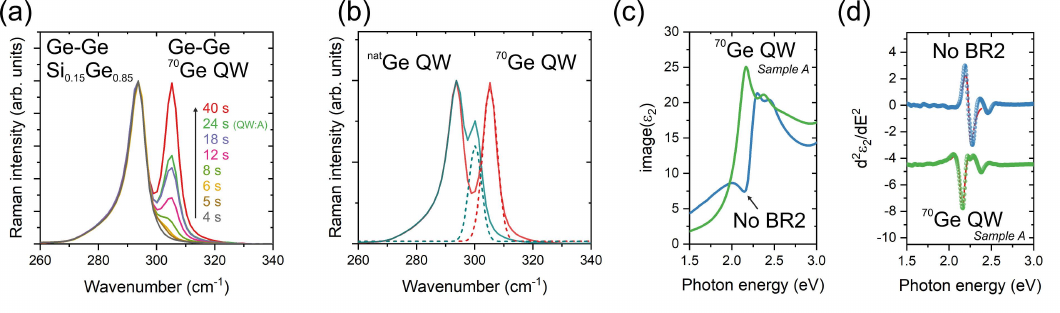}
    \caption{\textbf{Vibrational and optical properties of $^\text{70}$Ge QWs}. (a) Ge-Ge LO vibrational mode recorded for a set of $^\text{70}$Ge QW grown with a variable growth time between 4 and 40 s corresponding to a 3-30 nm thickness range. (b) Ge-Ge LO vibrational mode in two identical $^\text{nat}$Ge and $^\text{70}$Ge QW samples. (c) The imaginary dielectric function of an 18 nm-thick $^\text{70}$Ge QW structure (green) and for a reference material consisting of the same layer but without BR2 (blue). (d) CP analysis of the corresponding dielectric function. The CP energy position is extracted after fitting simultaneously the real and imaginary part of the second derivative of the dielectric function.}
    \end{figure*}

\begin{figure*}[htb]
    \centering
    \includegraphics[width=0.45 \textwidth]{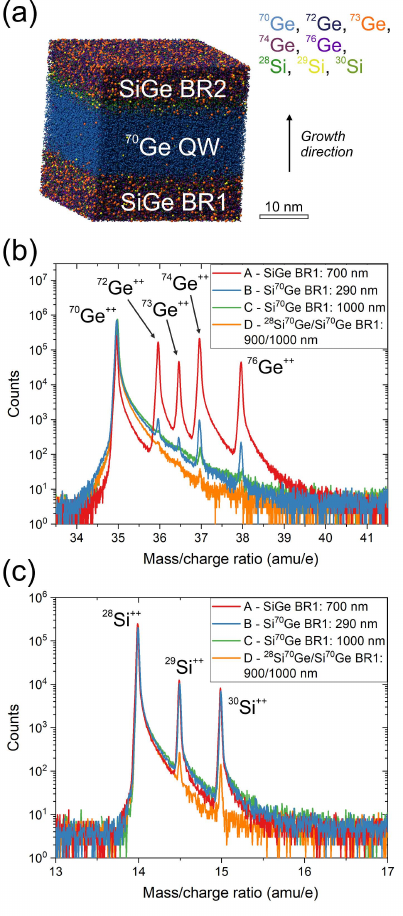}
    \caption{\textbf{Atom probe tomography of $^\text{70}$Ge QWs}. (a) 3D atom-by-atom APT map of a $^\text{70}$Ge/SiGe heterostructure. The 30 × 30 × 30 nm$^\text{3}$ map contains approximately 900,000 atoms. (b) Ge-Ge LO vibrational mode in two identical $^\text{nat}$Ge and $^\text{70}$Ge QW samples. Mass spectra of the investigated QW samples showing doubly charged Ge ions (b) and Si ions (c) of 10 million atoms.}
    \end{figure*}

A first glimpse into the isotopic content of the as-grown QWs was obtained from Raman spectroscopy studies. Fig. 2(a) shows Raman spectra around the Ge-Ge LO mode recorded for a set of QWs grown at a variable growth time between 4 and 40 s corresponding to a 3-30 nm thickness range. The spectra indicate the presence of two distinct modes. The first is centered around 293.7 cm$^\text{-1}$ corresponding to Ge-Ge LO mode in the SiGe barrier, whereas the second peak at 305.3 cm$^\text{-1}$ is attributed to the same mode but in the $^\text{70}$Ge QW. This assessment is consistent with the observed increase in the second peak intensity as the QW thickness increases. Note that the Ge-Ge mode in $^\text{nat}$Ge QW is detected at 300.1 cm$^\text{-1}$, as demonstrated in Fig. 2(b) comparing two identical $^\text{nat}$Ge and $^\text{70}$Ge QW samples. The observed shift between the two samples is analyzed based on the quasi-harmonic approximation, which is a valid approximation for semiconductors at room temperature ~\cite{MOUTANABBIR2009,Mukherjee2015}. According to the virtual crystal approximation, a simple harmonic analysis predicts that the energy of a phonon mode is inversely proportional to the square root of the average isotopic mass. The average isotopic mass is given by ⟨\emph{m}⟩ = $\Sigma$$_\text{i}$c$_\text{i}$\emph{m}$_\text{i}$, with c$_\text{i}$ being the fractional composition of an isotope of mass \emph{m}$_\text{i}$. Knowing that the atomic mass of $^\text{nat}$Ge is 72.63 amu, the measured wavenumbers of Ge-Ge LO mode in the sets of QWs yield an average atomic mass in $^\text{70}$Ge QW lattice of 70.17 amu corresponding to at least 99.6$\%$ enrichment in $^\text{70}$Ge isotopes. As discussed below, the growth protocol has a strong effect on the isotopic purity of the QW. 

It is important to mention that the limited spectral resolution (~1 cm$^\text{-1}$) of the used Raman setup does not allow addressing the effect of isotopic purification on lattice disorder ~\cite{Mukherjee2015}. Nevertheless, it is reasonable to conclude that the similarity observed in the full width at the half maximum of the Ge-Ge peaks in $^\text{nat}$Ge and $^\text{70}$Ge is indicative of a similar crystalline quality, which is consistent with XRD and TEM studies. To further assess the quality of the grown QWs, SE studies were carried out on $^\text{70}$Ge QW samples. For these studies, reference samples consisting of the same grown layers but without BR2 were also prepared and investigated. Fig. 2(c) displays the measured spectra for 18 nm $^\text{70}$Ge QW and the associated reference material. The figure shows the imaginary dielectric function (left) and the critical point (CP) analysis of the measured dielectric function. The nature of the lineshape of the dielectric function of both heterostructures conceals insights into the quantum confinement in the $^\text{70}$Ge QW. Note that the penetration depth of the incident excitation near the E$_\text{1}$ CP is around 20-35 nm for Ge bulk. If one considers a limited spectral range between 1.5-3 eV, the effect of the underlying materials (SiGe buffer, Ge-VS, and Si substrate) can be negated as the incident light will not reach and excite them. Consequently, only the top three layers ($^\text{70}$Ge QW, BR2, and Si cap) should in principle contribute to the measured dielectric function (Fig. 2(c)). Moreover, the contribution of the ~3-5 nm-thick Si cap should be excluded in the analysis as the E$_\text{1}$ CP of Si is located around 3.4 eV ~\cite{Lautenschlager1987}, which is outside the measured spectral range. 

The second derivative of the dielectric function of the two samples (with and without the top barrier BR2) is displayed in Fig. 2(d). To unravel the electronic structure of the analyzed heterostructure, the measured data were fitted using a generic critical point parabolic band model ~\cite{Aspnes1983,Lautenschlager1987,Vina1984,Assali2022}. The CP energy of the $^\text{70}$Ge layer without a barrier is evaluated at 2.156 eV, which is close to the Ge bulk CP of 2.134 eV ~\cite{Vina1984}, whereas for $^\text{70}$Ge QW sample a blueshift is noted yielding a CP energy of 2.233 eV. More importantly, the qualitative difference between both dielectric functions at 2.17 eV is clear. Indeed, the CP lineshape changes drastically from 2D Van Hove singularities in the reference structure (green dots) to a discrete excitonic lineshape in $^\text{70}$Ge QW (blue dots). This observed change in CP lineshape and energy is indicative of quantum confinement and its associated narrowing of the optical transition in Ge ~\cite{Assali2022}.

In the following, the isotopic content of the grown QWs is discussed based on APT studies. Fig. 3(a) shows a representative 3D 30 × 30 × 30 nm$^\text{3}$ atom-by-atom APT map of a $^\text{70}$Ge QW. The map indicates that the QW region contains mainly the $^\text{70}$Ge isotope, but traces of other isotopes can also be seen. Before quantifying and discussing the level of these contaminants, the recorded mass spectra are described first, as shown in Fig. 3(b,c). The figures exhibit the mass spectra recorded for a set of four QW samples labeled A, B, C, and D, as illustrated in Fig. 1(a). These samples were grown under different conditions. In sample A, the QW was grown using $^\text{70}$GeH$_\text{4}$, whereas the SiGe barriers were grown using $^\text{nat}$GeH$_\text{4}$. In the other three samples, the growth of both barriers and QWs was conducted using $^\text{70}$GeH$_\text{4}$. However, the change from $^\text{nat}$GeH$_\text{4}$ to $^\text{70}$GeH$_\text{4}$ occurred during the growth of the underlying SiGe layer at a variable thickness from the interface with the QW: 290 nm (B), 1000 nm (C), and 1890 nm (D). This means that the changes from $^\text{nat}$GeH$_\text{4}$ to $^\text{70}$GeH$_\text{4}$  took place at different times during the growth of SiGe buffer layer prior to the QW growth in these samples (B: 8 min, C: 24 min, and D: 29 min). In the case of sample D, the growth of SiGe barriers was conducted using the isotopically purified precursor $^\text{28}$SiH$_\text{4}$ instead of $^\text{nat}$Si$_\text{2}$H$_\text{6}$. The growth rate was higher for this sample due to a higher GeH$_\text{4}$ supply required for the growth optimization using $^\text{28}$SiH$_\text{4}$ precursor. The obtained APT mass spectra are compared in Fig. 3(b,c) showing the spectra of doubly charged Ge ions (Fig. 3(b)) and doubly charged Si ions (Fig. 3(c)). Each spectrum contains 10 million atoms from the selected region which includes most of the top barrier, the full QW and its interfaces, and a part of the bottom barrier. Note that this includes the QW interfaces and the local fluctuations in the isotopic purity observed near these interfaces, as shown in Fig. 4. 

The mass spectrum of sample A shows peaks associated with all five Ge isotopes at intensities close to the natural abundance of each isotope as most of the signal originates from the barriers grown with $^\text{nat}$GeH$_\text{4}$ (Fig. 3(b)). However, in samples B, C, and D, the APT spectra clearly show enrichment in $^\text{70}$Ge isotope as the peaks related to other isotopes have significantly diminished. Interestingly, the level of this contamination from other isotopes is intimately related to the growth protocol. Indeed, the level of Ge isotope cross-contamination becomes lower the longer the time, relative to the moment of the QW growth, of the transition from $^\text{nat}$GeH$_\text{4}$ to $^\text{70}$GeH$_\text{4}$. This indicates that the detection of $^\text{70}$Ge$^\text{++}$, $^\text{73}$Ge$^\text{++}$, $^\text{74}$Ge$^\text{++}$, and $^\text{76}$Ge$^\text{++}$ peaks is a manifestation of the reservoir effect, meaning that $^\text{nat}$GeH$_\text{4}$ used to grow the much thicker Ge-VS and SiGe-VS still resides in the growth reactor for an extended period of time. This leads to the undesired incorporation of the nuclear spin-full $^\text{73}$Ge isotope into the growing QW structure. Herein, it is shown that an early introduction of $^\text{nat}$GeH$_\text{4}$ can eliminate this contamination to a great extent. Ideally, the growth of the entire stack Ge-VS/SiGe-VS/BR1/Ge/BR2 should be done using $^\text{70}$GeH$_\text{4}$, but the process can be costly. Similarly, Fig. 3(c) shows that the use of $^\text{28}$SiH$_\text{4}$ to grow the SiGe barriers leads to a significant reduction, more than 30-fold, of the amount of $^\text{29}$Si isotope in the heterostructure. Since the hole wavefunction in Ge QW is expected to leak to the SiGe barriers, it is also important to suppress the hyperfine interactions that may result from the presence of $^\text{29}$Si isotope.

\begin{figure*}[htb]
    \centering
    \includegraphics[width=0.85 \textwidth]{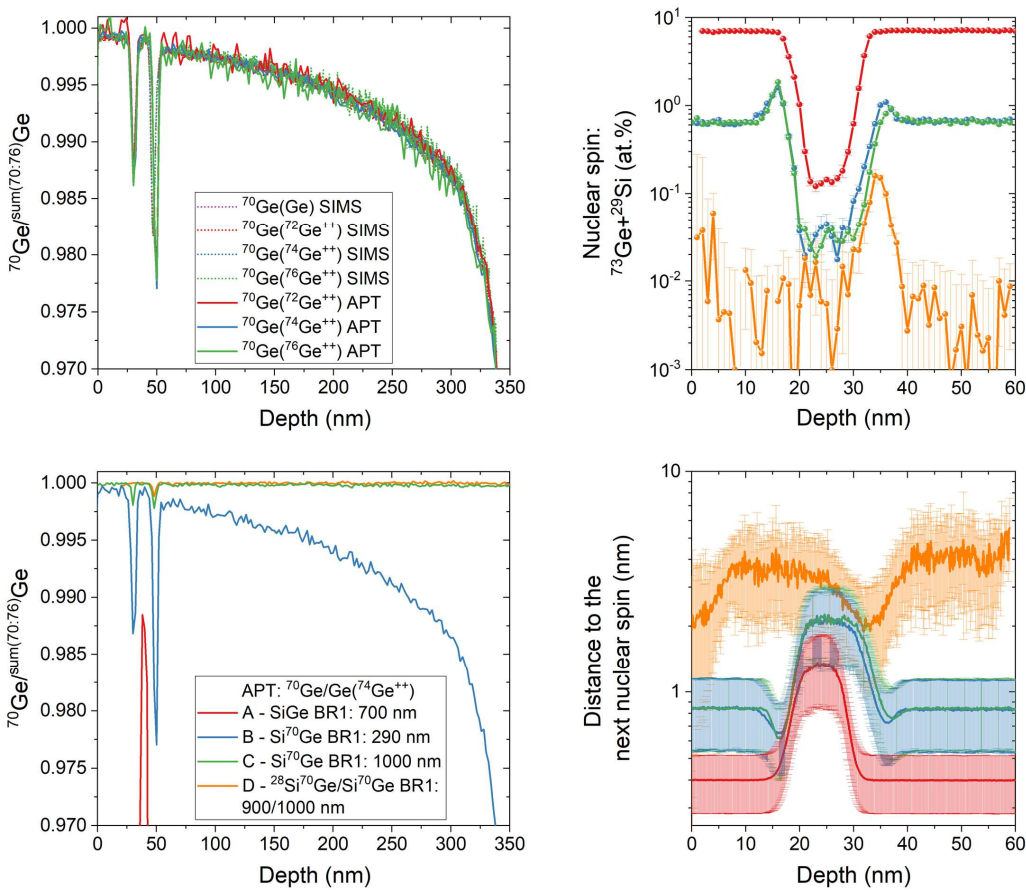}
    \caption{\textbf{Isotopic purity and nuclear spin background in $^\text{70}$Ge QWs}. (a) The evolution of $^\text{70}$Ge content across sample B as measured by SIMS (dotted lines) and APT with tail-corrected analysis (solid lines) from $^\text{72}$Ge, $^\text{74}$Ge, and $^\text{76}$Ge signals. (b) The depth profile of the isotopic purity across the investigated heterostructures. (c) The concentration profiles of nuclear spin-full species $^\text{73}$Ge and $^\text{29}$Si in the investigated heterostructures. (d) The evolution as a function of the thickness of the average distance between nuclear spins in the investigated heterostructures.}
    \end{figure*}

The local isotopic purity and 3D distribution of isotopes can be obtained from APT. However, since the peaks of heavier isotopes are embedded in the tails of the lighter isotopes (Fig. 3(b)), it is important to carefully analyze and model the mass spectra to separate the tails and the peaks to accurately quantify the isotopic content in the heterostructures. Herein, SIMS analyses were carried out to validate the APT isotope mapping method. Since all non-$^\text{70}$Ge isotopes originate from a natural Ge source, one can use the content measured for each isotope to estimate the $^\text{70}$Ge purity by making a projection of the overall contamination based on the natural distribution of isotopes. As shown in Fig. 4(a), SIMS data provide estimates derived from $^\text{72}$Ge, $^\text{74}$Ge, and $^\text{76}$Ge signals coinciding almost perfectly which each other and the estimate for the  $^\text{70}$Ge purity gained from considering the signal from all Ge isotopes. For APT, however, a difference was observed (data not shown) when estimating based on doubly charged $^\text{72}$Ge, $^\text{74}$Ge, or all of the isotopes. This discrepancy is caused by the aforementioned overlap of isotope peaks in the mass spectra (Fig. 3(b)). To address this issue, a Monte-Carlo approach is implemented where the tails are fitted locally around the peak region, and the peak and tail are decomposed multiple 100s or 1000s times to find the average content of the peak and the error created by the decomposition. The resulting estimate using the tail-corrected data for doubly-charge $^\text{72}$Ge,  $^\text{74}$Ge, and  $^\text{76}$Ge match SIMS data, as shown in Fig. 4a (solid line). 

Using the same Monte Carlo approach, we can quantify the $^\text{70}$Ge purity in all samples. The result is shown in Fig. 4(b) highlighting once more the differences between the samples in terms of isotopic purity near the QW caused by the difference in time passed between the onset of $^\text{70}$GeH$_\text{4}$ growth and the QW growth. Furthermore, both SIMS and APT data consistently show that the 90 s growth interruption at the QW interfaces, introduced to promote the growth of sharper interfaces, leads to an accumulation in $^\text{nat}$Ge at the interface. For the growth of sample D, the top barrier was grown without interruption thus suppressing the isotopic cross-contamination at the interface. Maintaining the $^\text{70}$Ge purity is important to achieve a nuclear spin-depleted interface and BR1. 

A more accurate evaluation of the nuclear spin background is obtained from APT analyses displayed in Fig. 4(c). The figure outlines the total concentration profiles of nuclear spin-full isotopes $^\text{73}$Ge and $^\text{29}$Si across the investigated heterostructures. It is noticeable that in $^\text{nat}$Si$^\text{nat}$Ge/$^\text{70}$Ge/$^\text{nat}$Si$^\text{nat}$Ge (sample A) the nuclear spin concentration drops from 6 at.$\%$ in the SiGe barriers down to 0.1 at.$\%$ in the QW. This background is further reduced to 0.02 at.$\%$ in QWs of samples B and C and even below 0.01 at.$\%$ in sample D consisting of $^\text{28}$Si$^\text{70}$Ge/$^\text{70}$Ge/$^\text{28}$Si$^\text{70}$Ge. Besides providing the isotopic composition profiles, APT also allows extracting the atomic-level spatial distribution of individual nuclear spin-full species $^\text{73}$Ge and $^\text{29}$Si, as displayed in Fig. 4(d), The figure shows the depth evolution of the average distance between neighboring nuclear spins across the investigated heterostructures. To obtain these profiles, a model of SiGe lattice was generated from APT maps ~\cite{Wuetz2022} on which the distribution of each isotope was imprinted thus allowing the calculation of the distance between nuclear spins in a lattice plane-by-lattice plane fashion. The uncertainty in these calculations was assessed by sampling 10 different models. The obtained result demonstrates that the average distance between nuclear spins is the lowest in the QW for all samples, but it remains sensitive to the growth conditions. For instance, in $^\text{nat}$Si$^\text{nat}$Ge barriers (A) the obtained average distance is 0.3-0.4 nm, whereas it increases by one order of magnitude to 3-4 nm in isotopically pure$^\text{28}$Si$^\text{70}$Ge/$^\text{70}$Ge/$^\text{28}$Si$^\text{70}$Ge heterostructure (D).

\section{Conclusion}
In summary, this work demonstrates the epitaxial growth of nuclear spin-depleted, isotopically enriched $^\text{70}$Ge QWs. The growth was achieved on Si wafers using enriched precursors $^\text{70}$GeH$_\text{4}$ and $^\text{28}$SiH$_\text{4}$ in a reduced-pressure CVD system. The crystalline quality of the grown heterostructures was confirmed by XRD and electron microscopy studies. The critical point of the grown QWs exhibits a discrete excitonic lineshape at 2.233 eV indicative of quantum confinement. The isotopic purity and the distribution of the nuclear spin background were investigated using APT. In this regard, a Monte Carlo approach was introduced to solve the discrepancies in APT analyses caused by the overlap of isotope peaks in the recorded mass spectra. These analyses demonstrate that the isotopic content is very sensitive to the growth conditions including any growth interruption. The latter was found to induce an accumulation of natural Ge isotopes at the growth interface leading to lower $^\text{70}$Ge content. To evaluate the distribution of the residual nuclear spin background, a lattice model was constructed to map the average distance between the two nuclear spin-full isotopes $^\text{73}$Ge and $^\text{29}$Si. These studies showed that the distance between nuclear spins reaches 3-4 nm in $^\text{70}$Ge/$^\text{28}$Si$^\text{70}$Ge, which is an order of magnitude higher than in natural Ge/SiGe heterostructure. Additionally, the lowest concentration of $^\text{73}$Ge and $^\text{29}$Si contaminants in the heterostructure was established at 0.01$\%$ in both QW and barriers of $^\text{70}$Ge/$^\text{28}$Si$^\text{70}$Ge heterostructure. These insights constitute a valuable input to improve the design and theoretical modeling of spin qubits by providing quantitative, atomic-level details on nuclear spin distribution. 

\bigskip
\noindent {\bf METHODS}.

\noindent X-ray diffraction (XRD) measurements performed using a Bruker Discover D8. A 3 bounces Ge(220) 2-crystals analyzer was placed in front of the XRD detector during the XRD (004) and (224) reciprocal space map (RSM) analysis. The microstructure of the grown materials was investigated by transmission electron microscopy (TEM). TEM specimens were prepared in a Thermo Fisher Helios Nanolab 660 dual-beam scanning electron microscope using a gallium-focused ion beam (FIB) at 30, 16, and 5 kV. Electron beam-induced carbon and platinum were locally deposited on the sample to protect the imaged region from being damaged by the ion-beam milling during the thinning of the TEM lamella. TEM and scanning TEM (STEM) analyses were carried out on a Thermo Scientific Talos F200X S/TEM system with an acceleration voltage of 200 kV. 

Insights into the quality of interfaces, the atomic-level composition, and the isotopic purity were obtained using atom probe tomography (APT). APT specimens were prepared in a FEI Helios Nanolab 660 dual-beam scanning electron microscope using a gallium-focused ion beam (FIB) at 30, 16, and 5 kV. A 120-150 nm-thick chromium capping layer was deposited on the samples before FIB irradiation to minimize the implantation of gallium ions into the imaged region. APT studies were performed in a LEAP 5000XS tool. The LEAP 5000XS utilizes a picosecond laser to generate pulses at a wavelength of 355 nm. For the analysis, all samples were cooled to a temperature of 25 K. The experimental data were collected at laser powers of 3-6 pJ. Additional insights into the chemical and isotopic compositions are also obtained using secondary ion mass spectrometry (SIMS).

Raman scattering analyses were performed at room temperature using a 633 nm excitation laser. Additionally, the uniformity of the growth thickness as well as the optical signature of quantum confinement were investigated using spectroscopic ellipsometry (SE). SE measurements were carried out at room temperature, using a variable angle spectroscopic RC2-XI ellipsometer manufactured by J. A. Woollam Co. The variable angle spectroscopic ellipsometer system covers the 0.5–6 eV range. All heterostructures were measured between 70° and 80° angles of incidence with a 1° step. A noticeable increase in the sensitivity of the SE parameters ($\Psi$ and $\Delta$) was observed around 76-77°, which is very close to the Brewster angle for Si and Ge. Thus, during the optical modeling, special care was accorded to the modeling near this angle.

\bigskip
\noindent {\bf ACKNOWLEDGEMENTS}.
The authors thank J. Bouchard for the technical support with the CVD system. O.M. acknowledges support from NSERC Canada (Discovery Grants, Alliance International Quantum, and CQS2Q Consortium), Canada Research Chairs, Canada Foundation for Innovation, Mitacs, PRIMA Qu\'ebec, and Defense Canada (Innovation for Defense Excellence and Security, IDEaS), the European Union's Horizon Europe research and innovation programme under grant agreement No 101070700 (MIRAQLS), and the US Army Research Office Grant No. W911NF-22-1-0277.\\


\nocite{*}

\bibliography{main}

\end{document}